\documentclass{jltp}
\usepackage{amssymb}
\usepackage{graphicx}
\usepackage{bm}
\bmdefine\bomega{\omega} \bmdefine\bOmega{\Omega}
\bmdefine\bnabla{\nabla} \bmdefine\bkappa{\kappa}
\bmdefine\bphi{\phi}

\title{Transition to Superfluid Turbulence}

\author{V.B. Eltsov$^{*\dag}$, M. Krusius$^*$, and G.E. Volovik$^{*\ddag}$
\address{$^*$Low Temperature
Laboratory,  Helsinki University of Technology\\
P.O.Box 2200, FIN-02015 HUT, Finland\\
$^{\dag}$Kapitza Institute of Physical Problems, Kosygina 2,
119334 Moscow, Russia\\ $^{\ddag}$Landau Institute for Theoretical
Physics, Kosygina 2, 119334 Moscow, Russia }}

\runninghead{V.B. Eltsov, M. Krusius, and G.E. Volovik}{Transition
to Superfluid Turbulence}

\begin{document}

\maketitle \vspace{-6mm}

\begin{abstract}
Turbulence in superfluids depends crucially on the dissipative
damping in vortex motion. This is observed in the B phase of
superfluid $^3$He where the dynamics of quantized vortices changes
radically in character as a function of temperature. An abrupt
transition to turbulence is the most peculiar consequence. As
distinct from viscous hydrodynamics, this transition to turbulence
is not governed by the velocity-dependent Reynolds number, but by
a velocity-independent dimensionless parameter $1/q$ which depends
only on the temperature-dependent mutual friction -- the
dissipation which sets in when vortices move with respect to the
normal excitations of the liquid. At large friction and small
values of $1/q \lesssim 1$ the dynamics is vortex number
conserving, while at low friction and large $1/q \gtrsim 1$
vortices are easily destabilized and proliferate in number. A new
measuring technique was employed to identify this hydrodynamic
transition: the injection of a tight bundle of many small vortex
loops in applied vortex-free flow at relatively high velocities.
These vortices are ejected from a vortex sheet covering the AB
interface when a two-phase sample of $^3$He-A and $^3$He-B is set
in rotation and the interface becomes unstable at a critical
rotation velocity, triggered by the superfluid Kelvin-Helmholtz
instability.

%\textbf{\textit{KEY WORDS}}: quantized vortex; vortex dynamics;
%transition to turbulence; turbulence; mutual friction;
%Kelvin-Helmholtz instability
\vspace{-3mm}
\end{abstract}

%Include this space if you do not use sections in your document.
%\vspace{0.3in}

\section{INTRODUCTION}

Superfluid turbulence -- the tangled motion of quantized vortex
lines in superfluid $^4$He-II -- has been known to exist for fifty
years.\cite{Feynman} By increasing the applied flow velocity
beyond some relatively low critical value, at which vortices
become mobile, their turbulent motion is started and superfluid
flow becomes dissipative.\cite{Vinen} In this context we mean with
applied flow the {\em counterflow velocity} ${\bm v} = {\bm
v}_{\rm n} - {\bm v}_{\rm s}$, the difference between the
velocities ${\bm v}_{\rm n}$ of the normal and ${\bm v}_{\rm s}$
of the superfluid components, which is created by external means.

Only recently it has been realized from measurements on superfluid
$^3$He-B that turbulence is not necessarily a generic property of
all superfluids, but one whose presence crucially depends on the
damping in vortex motion.\cite{Nature} The damping, or more
accurately the mutual friction dissipation, turns out to govern
the onset of turbulence, dividing vortex motion in $^3$He-B to
superconductor-like regular behavior at high temperatures, where
the vortex number is conserved in dynamic processes, and to
$^4$He-like disordered behavior at low temperatures, where
turbulence easily sets in when perturbations are introduced in
superfluid flow.

The transition to turbulence in superfluids can be compared to
that in viscous flow, as discussed in Ref.~[\onlinecite{Mullin}]
on linear pipe flow with circular cross section. Here the
transition is governed by the velocity-dependent Reynolds number
$Re=UD/\nu$, where $U$ is the mean velocity, $D$ the
characteristic length scale (the pipe diameter in
Ref.~[\onlinecite{Mullin}]), and $\nu$ the kinematic viscosity. In
the 16\,m long circular pipe of Ref.~[\onlinecite{Mullin}] laminar
flow is stable at flow velocities $Re < 20\,000$. However, on
injecting a controlled square pulse of perturbing flow azimuthally
in the linear stream, the flow can be converted from laminar to
turbulent over some length of the pipe. This turbulence travels
downstream in the pipe, while upstream in the absence of the
perturbation laminar flow again recovers. The critical amplitude
of the perturbing mass flux $\Phi_{\rm inj}$, required to reach
the transient turbulent state, was found to obey a scaling law of
the form $\Phi_{\rm inj}/\Phi_{\rm pipe} = Re^{-\delta}$, where
the exponent has the value $\delta = 1 \pm 0.01$.

To generate the transition to turbulence in superfluids, one or
several vortices are injected in rotating vortex-free flow of
$^3$He-B.\cite{Turbulence} It is then found that the transition is
not governed by viscosity (which is absent for the superfluid
fraction of the liquid), but by the mutual friction between
vortices and the normal fraction of the liquid. As distinct from
viscosity $\eta$, which enters the Reynolds number as the
dimensional kinematic viscosity $\nu = \eta / \rho$, mutual
friction is described by two dimensionless parameters which
represent its dissipative and reactive components. In $^3$He-B in
the range of the transition to turbulence both parameters are of
comparable magnitude and have to be taken into account. In
$^4$He-II the reactive mutual friction parameter is much smaller
and is usually neglected (see review
[\onlinecite{VinenIntroduction}]). In the superfluid, the
viscosity dependent Reynolds number has to be replaced by a
dimensionless characteristic number which is called $1/q$ and only
depends on the two mutual friction parameters. In particular as
opposed to Reynolds number, $1/q$ is velocity independent which is
also observed in measurement at higher flow velocities as the
limiting case. In this short review we focus on this limiting
regime, which corresponds to the case when a sufficient number of
closely spaced seed vortex loops is injected, so that they
immediately start interacting and instantaneously produce
turbulence.\cite{Nature}

If the number of seed vortices is reduced or their spacing is
increased, then the flow perturbation is weakened and the
transition to turbulence moves to higher values of $1/q$. In this
case more new vortices need to be generated, before they can start
interacting turbulently. Measurements with injection down to the
limit of one single seed vortex are discussed in
Refs.~[\onlinecite{Neutron,Solntsev}]. Obviously such cases
require some additional mechanism, which leads to an increase in
the number of vortices in the low-density regime, when vortices do
not yet interact. This is the single vortex instability in applied
flow which via loop formation and reconnection generates new
independent vortices. In this way the transition to turbulence
becomes a complex process of series coupled mechanisms in the
regime of small flow perturbation.\cite{Precursor} However,
independently of the applied perturbation, in all these
measurements the onset of turbulence is displayed as an abrupt
transition, which takes place within a narrow distribution of
$1/q$ values. Moreover, the average of this distribution of $1/q$
values proves to depend on the magnitude of the applied
perturbation in a power-law manner.

Comparing measurements on the transition to turbulence in viscous
and superfluid flow, we notice that they proceed in somewhat
different manner. Nevertheless, there are similarities: In both
cases (i) the initial state is perturbed externally by means of a
quantitatively controlled disturbance which (ii) sets off
turbulence for a short length of time if (iii) the perturbation is
of sufficient amplitude, with power-law dependence on the relevant
controlling parameter (which is $Re$ in viscous flow and the
mutual friction dependent parameter $1/q$ in superfluids).

There are special reasons why the transition to turbulence as a
function of mutual friction has not been observed in superfluid
$^4$He-II and was only recently discovered in superfluid $^3$He-B.
As distinct from $^4$He-II, $^3$He-B is a Fermi superfluid, where
the superfluidity is caused by Cooper pairing. The mutual friction
between vortices and the normal fraction of the liquid, which is
composed of fermionic quasiparticles, is mediated by
quasiparticles populating the vortex core states,\cite{KopninBook}
the so-called fermion zero modes\cite{VolovikBook}. The scattering
between the two types of quasiparticles leads to mutual friction
and is described by a theory similar to the BCS theory of
superconductivity. As a result the parameter $q(T)$ appears to be
a dimensionless function of the dimensionless parameter $T/T_{\rm
c}$. In Fermi superfluids in the weak coupling approximation this
parameter crosses unity at $T\sim 0.6 T_{\rm c}$, {\it i.e.} in
the middle of the experimentally accessible temperature range of
$^3$He-B. (In the cold superfluid fermionic gases discussed in
Ref.~[\onlinecite{Combescot}] $q$ can be adjusted with a magnetic
field if the system is close to the Feshbach resonance.)

In contrast, in the boson superfluid $^4$He-II vortex dynamics is
practically always in the turbulent regime (see review
[\onlinecite{VinenIntroduction}]). Regular flow of vortices could
be perhaps expected only within microkelvins from the superfluid
transition temperature $T_\lambda$, but there is not yet enough
information on vortex dynamics in this regime. Even there, the low
viscosity of the normal component (in $^4$He-II the normal
component is one of the least viscous fluids existing) causes its
flow to become easily turbulent, which can in turn influence the
flow of the superfluid component. In contrast, in $^3$He-B the
normal component has $\sim 10^4$ times higher oil-like viscosity
and is practically always in a state of laminar flow. The absence
of turbulence in the flow of the normal component of $^3$He-B
amounts to a considerable simplification and leads to new effects,
which are absent in $^4$He-II. An example is a new scaling law for
the Kolmogorov-Richardson cascade in developed homogeneous
superfluid turbulence.\cite{LNV,LNS}

The injection mechanism, which led to the discovery of the
transition to turbulence as a function of $1/q$, is of particular
interest. Here the injected seed vortices originate from the AB
interface in a two-phase sample of $^3$He-A and $^3$He-B. The seed
vortices are tossed as a tight bundle of some 10 loops across the
AB interface from $^3$He-A into the vortex-free flow of $^3$He-B.
This happens when the interface becomes unstable with respect to
wave formation at a well-defined critical value for superfluid
counterflow parallel to the interface. During the non-linear stage
of this corrugation instability, the vortices in the deepest
corrugation of the interface wave are ejected on the B-phase side
of the interface. The instability itself, known as the superfluid
Kelvin-Helmholtz shear flow instability, is reproducible and
predictable, its measurements and theory match without fitting
parameters. This is different from the ordinary Kelvin-Helmholtz
instability at the interface between two viscous liquids or gases.
In viscous fluids the initial state is not well described, since
the shear-flow configuration is not an equilibrium situation and
cannot be expressed as a solution of the Navier-Stokes equation.
Interestingly, the superfluid Kelvin-Helmholtz instability shares
some characteristics with the instability of quantum vacuum within
the horizon or ergoregion of the black hole.\cite{BHandWH}

In this short review\cite{ROPreview} we shall first introduce the
superfluid Kelvin-Helmholtz instability and the injection of
vortex seed loops in Sec. 2. Then follows in Sec. 3 a description
of the transition to turbulence in the case when turbulence is
instantaneously started by the injected vortex loops.
\vspace{-5mm}

\section{SUPERFLUID KELVIN-HELMHOLTZ INSTABILITY }

\label{experiment}

\subsection{Kelvin-Helmholtz Instability in Viscous Liquids}
\label{InterfaceInstability}

Kelvin--Helmholtz (KH) instability is one of the many interfacial
instabilities in the hydrodynamics of liquids, gases, charged
plasma, and even granular materials. It refers to the dynamic
instability of an interface with discontinuous tangential flow
velocities and can loosely be defined as the instability of a
vortex sheet. Many natural phenomena have been attributed to this
instability. The most familiar ones are the generation of
capillary waves on the surface of water, first analyzed by Lord
Kelvin,\cite{LordKelvin} and the flapping of sails and flags,
first discussed by Lord Rayleigh.\cite{Rayleigh}

Many of the leading ideas in the theory of interfacial
instabilities in hydrodynamics were originally inspired by
considerations about ideal inviscid flow. A horizontal interface
between two ideal liquids, stacked on top of each other by gravity
because of their different mass densities $\rho_1$ and $\rho_2$,
and flow parallel to the interface at velocities ${\bf v}_1$ and
${\bf v}_1$, leads to a corrugation instability at the critical
differential flow velocity\cite{landau_fluid_dynamics}
\begin{equation}
({\bf v}_1-{\bf v}_2)^4= 4\sigma g (\rho_1-\rho_2) \, \frac{
(\rho_1+\rho_2)^2}{ \rho_1^2\rho_2^2}~. \label{KHClassical}
\end{equation}
Here $\sigma$ is the surface tension of the interface and $g$
gravitational acceleration. To separate the gravitational and
inertial properties of the liquids, let us rewrite the threshold
velocity in the following form
\begin{equation}
  {\rho_1\rho_2\over \rho_1+\rho_2} \, ({\bf v}_1-{\bf v}_2)^2=2\sqrt{\sigma
F}~. \label{InstabilityCondition1}
\end{equation}
We associate $F$ with the external field stabilizing the position
of the interface, which in the gravitational field is the gravity
force
\begin{equation}
F=g(\rho_1-\rho_2)~, \label{GravityForce}
\end{equation}
but which in the general case can originate from some other
source. The surface mode of ripplons or capillary waves, which is
first excited at the instability, has the wave number
corresponding to the inverse `capillary length',
\begin{equation}
k_0=\sqrt{F/\sigma}~. \label{WaveVectorInstability}
\end{equation}
However, ordinary fluids are not ideal and the correspondence
between this theory and experiment is not good. One reason for
this is that one cannot properly prepare the initial state -- the
shear-flow discontinuity is never in equilibrium in a viscous
fluid. It is not a solution of the Navier-Stokes equation. That is
why one cannot properly extend the `instability' of the inviscid
case to finite viscosities. \vspace{-5mm}

\subsection{Kelvin-Helmholtz Instability in Superfluids}
\label{kh_superfluid}

In superfluids the criterion for the instability can be formulated
in the absence of viscosity, since the tangential velocity
discontinuity at the interface between $^3$He-A and $^3$He-B is a
stable non-dissipative state. These two superfluid phases have
different magnetic properties and their interface is stabilized by
the gradient in the applied magnetic field $H(z)$ which provides
the restoring force $F$ in Eq.~(\ref{InstabilityCondition1}):
\begin{equation}
F= {1\over 2}\nabla \left((\chi_{\rm A} -\chi_{\rm B} ) H^2\right)
~. \label{InstabilityCondition2}
\end{equation}
Here $\chi_{\rm A}>\chi_{\rm B}$ are the magnetic susceptibilities
of the A and B phases, respectively. One might expect that by
substituting this interfacial restoring force $F$ into
Eq.~(\ref{InstabilityCondition1}) and using the superfluid
densities of the A and B phases instead of the total density, one
obtains the critical velocity for the KH instability of the A-B
interface. However, it turns out that a proper extension of the KH
instability to superfluids incorporates the criterion in
Eq.~(\ref{InstabilityCondition1}) only as a particular limiting
case.

The criterion for the KH instability of ideal fluids in
Eq.~(\ref{InstabilityCondition1}) depends only on the relative
velocity across the interface. In practice there always exists a
preferred reference frame, imposed by the environment. In the
superfluid case it is the frame of the normal component moving
with velocity $\mathbf{v}_\mathrm{n}$ (in equilibrium
$\mathbf{v}_\mathrm{n}=0$ in a frame frame fixed to the rotating
container). Owing to this interaction of the AB interface with its
environment, the instability occurs at a lower differential flow
velocity than the classical criterion in
Eq.~(\ref{InstabilityCondition1}) assumes (see
Refs.~[\onlinecite{Volovik,BHandWH,Abanin}]):
\begin{equation}
{1\over 2} \rho_{\mathrm{s}B}({\bf v}_{\rm B} -{\bf v}_{\rm n})^2
+ {1\over 2} \rho_{\mathrm{s}A}({\bf v}_{\rm A} -{\bf v}_{\rm
n})^2 =\sqrt{\sigma F}~. \label{InstabilityConditionNewnon-zeroT}
\end{equation}
Here ${\bf v}_{\rm A}$ and ${\bf v}_{\rm B}$ are the velocities of
the superfluid components on the A- and B-phase sides of the
interface; while $\rho_{\mathrm{s}A}$ and $\rho_{\mathrm{s}B}$ are
the corresponding densities of these superfluid fractions.
\vspace{-5mm}

\subsection{Observation of Kelvin-Helmholtz Instability}
\label{ObservSuperfluid}

%%%%%%%%%%%%%%%%%%%%%%%%%%%%%%%%%%%%%%%%%%%%%%%%%%%%%%%%%%%%%%%%%%%%%%%
\begin{figure}[t]
\centerline{\includegraphics[width=\linewidth]{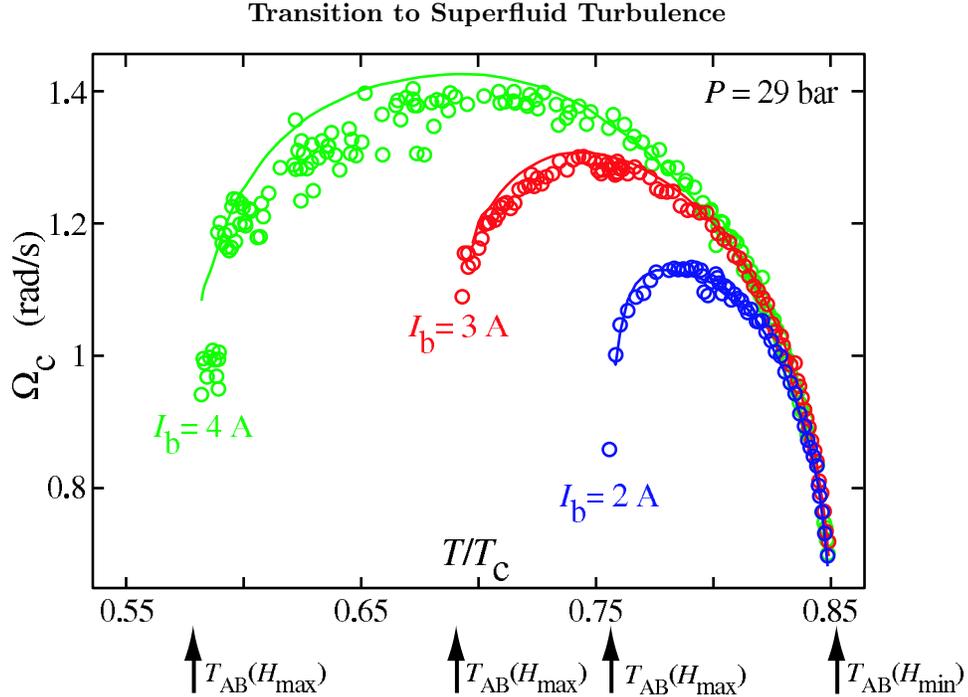}}
\medskip
\caption{Critical rotation velocity $\Omega_{\rm c} \approx
(v_{\rm n} -v_{\rm B})/R$ of the AB-interface instability versus
temperature at different currents $I_{\rm b}$ in the barrier
magnet which is employed to stabilize the A phase. If the barrier
field $H(z)$ exceeds the critical value $H_{\rm AB}(T,P)$ needed
for the A phase, then the AB interface resides at the two
locations $z$ where $H(z) = H_{\rm AB}(T,P)$. These locations and
the magnetic restoring force in Eq.~(\ref{InstabilityCondition2})
depend on temperature. The temperatures, where the AB interface
disappears, are indicated by vertical arrows below the figure for
different values of $I_{\rm b}$. The solid curves represent the
instability criterion
(\protect\ref{InstabilityConditionNewnon-zeroT}).}
\label{KHInstabilityCurvesFig}
\end{figure}
%\vspace{-5mm}
%%%%%%%%%%%%%%%%%%%%%%%%%%%%%%%%%%%%%%%%%%%%%%%%%%%%%%%%%%%%%%%%%%%%%%%

A comparison of Eq.~(\ref{InstabilityConditionNewnon-zeroT}) to
the measured critical rotation velocity $\Omega_{\rm c}$ of the
first KH instability event is shown in
Fig.~\ref{KHInstabilityCurvesFig}.\cite{Kelvin-HelmholtzInstabilitySuperfluids}
Here we set $v_{\rm B}=\Omega R$ and $v_{\rm A}= v_{\rm n}=0$
({\it i.e.} the velocities are given in the frame of the rotating
container, and $R$ is the radius of the container). No fitting
parameters are used. The curves have been drawn using generally
accepted values for the different superfluid $^3$He parameters.
Even the remaining residual differences between measurement and
Eq.~(\protect\ref{InstabilityConditionNewnon-zeroT}) have
plausible explanations which are discussed in
Ref.~[\protect\onlinecite{Kopu}]. Such remarkable agreement for a
complicated phenomenon can only be achieved in superfluids, where
shear flow is not dissipative until the instability threshold is
reached. Although the instability in
Eq.~(\ref{InstabilityConditionNewnon-zeroT}) depends on the
reference frame fixed to the normal component, this does not mean
that the renormalized instability criterion would depend on the
magnitude of the interaction with the normal component -- in fact,
it is still determined by only thermodynamics. Waves are formed on
the interface when the free energy of a corrugation becomes
negative in the reference frame of the environment.

In terms of the jargon accepted in general relativity, the
so-called ergoregion is then formed -- the region in which the
(free) energy of some excitations is negative. This connection
between the interface instability, horizons, and ergoregions in
black hole physics is worked out in
Refs.~[\onlinecite{BHandWH,SchutzholdUnruh}]. In the ergoregion
the AB interface becomes thermodynamically unstable, since it
becomes possible to reduce the energy via the normal component and
its interaction with the solid sample boundary. The original
classical instability condition Eq.~(\ref{InstabilityCondition1})
of the ideal inviscid fluid is restored if the interaction with
the environment is not effective. This might occur in the
superfluid, for example, during rapid rotational acceleration at
very low temperatures when the instability caused by the
interaction with the environment has not had enough time to
develop.

However, measurements now attest that down to moderately low
temperatures of $0.35\,T_{\rm c}$ the instability condition is not
given by the classical KH expression (\ref{InstabilityCondition1})
even in a perfectly inviscid superfluid, but by the renormalized
criterion in Eq.~(\ref{InstabilityConditionNewnon-zeroT}). Its
central property is that the instability condition does not depend
on the relative velocity of the two superfluids, but on the
velocity of each of the superfluids with respect to the
environment. The instability will occur even if the two fluids
have equal densities, $\rho_{\rm A} =\rho_{\rm B}$, and move with
the same velocity, ${\bf v}_{\rm A}={\bf v}_{\rm B}$. The
instability also occurs if there is only a single superfluid with
a free surface. These new features arise from the two-fluid nature
of the superfluid. The situation resembles that of a flag flapping
in wind, which was originally discussed in terms of the KH
instability of ideal inviscid fluids by Lord
Rayleigh.\cite{Rayleigh} Newer explanations involve boundary
interactions between the flag and the two gas streams which become
turbulent within a narrow boundary layer. The superfluid analogue
is an instability of a passive deformable membrane between two
distinct parallel streams having the same density and velocity.
The flag is the AB interface and the flagpole, which pins the
flag, serves as the reference frame fixed to the environment so
that Galilean invariance is violated. %\vspace{-3mm}

%%%%%%%%%%%%%%%%%%%%%%%%%%%%%%%%%%%%%%%%%%%%%%%%%%%%%%%%%%%%%%%%%%%%
\begin{figure}[t]
\centerline{\includegraphics[width=0.9\linewidth]{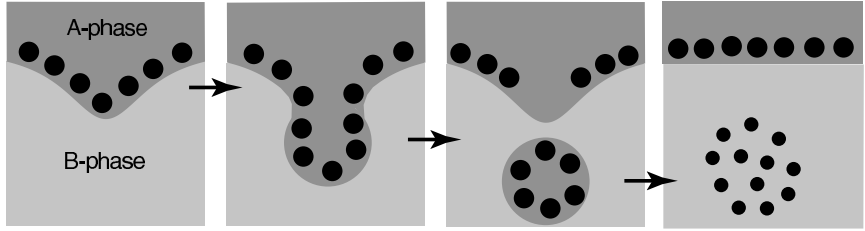}}
\medskip
\caption{One scenario for the transfer of circulation across the
AB interface in a KH instability. The A-phase vorticity is
confined by the Magnus lift force to a vortex layer which covers
the AB interface.\protect\cite{Hanninen} The Magnus force arises
from the tangential B-phase superflow below the interface. The
instability creates a corrugation in the interface which becomes a
potential well for A-phase vortices there. This pushes the
vortices deeper in the potential well, deforming the corrugation
to a trench with a droplet-like cross section. The trench, filled
with A-phase vorticity, then propagates in the bulk B-phase where
A phase is unstable and the multiply-quantized vortex relaxes to
singly-quantized B-phase vortices. In this scenario the number of
vortices ejected across the interface corresponds to the number of
vortices in the A-phase vortex layer within one corrugation or one
half wave-length of the interface `ripplon', which is in
accordance with
measurements.\protect\cite{Kelvin-HelmholtzInstabilitySuperfluids}
} \label{KelvinInstabilityFig}
\end{figure}
\vspace{-7mm}
%%%%%%%%%%%%%%%%%%%%%%%%%%%%%%%%%%%%%%%%%%%%%%%%%%%%%%%%%%%%%%%%%%%%

\subsection{Vortex Injection}
\label{VortexInjection}

The non-linear stage of the interface instability, when the
interface is distorted following the break-down of its ultimate
equilibrium condition, leads to the injection of a bundle of
vortices from the AB interface in the rapidly flowing B-phase,
when viewed in the frame of the rotating sample container. The
mechanism of injection has not been worked out in detail, but a
simplified scenario is presented in
Fig.~\ref{KelvinInstabilityFig}. In general, this mechanism can be
associated with the non-linear development of the vortex sheet
instability, as discussed {\it e.g.} in
Ref.~[\onlinecite{AbidVerga}].

In this context the detailed properties of the injection event are
not all that important. Rather we are interested in the dynamic
evolution of this bundle of vortices which initially (following
the instability) protrudes out of the AB interface and then
continues to the cylindrical sample boundary in the B-phase
volume. The measurements display at first glance a rather
unexpected result: The final B-phase state, {\it i.e.} the state
which after some transient development will be stable in time (at
constant rotation, temperature, and pressure), does not depend on
the initial velocity of B-phase flow, but only on temperature. %\vspace{-5mm}

\section{SUPERFLUID TURBULENCE}
\label{SuperfluidTurb}

\subsection{Transition to Turbulence as a Function of Mutual Friction}
\label{Exp}

%%%%%%%%%%%%%%%%%%%%%%%%%%%%%%%%%%%%%%%%%%%%%%%%%%%%%%%%%%%%%%%%%
\begin{figure}[t]
\begin{center}
\includegraphics[width=0.9\linewidth]{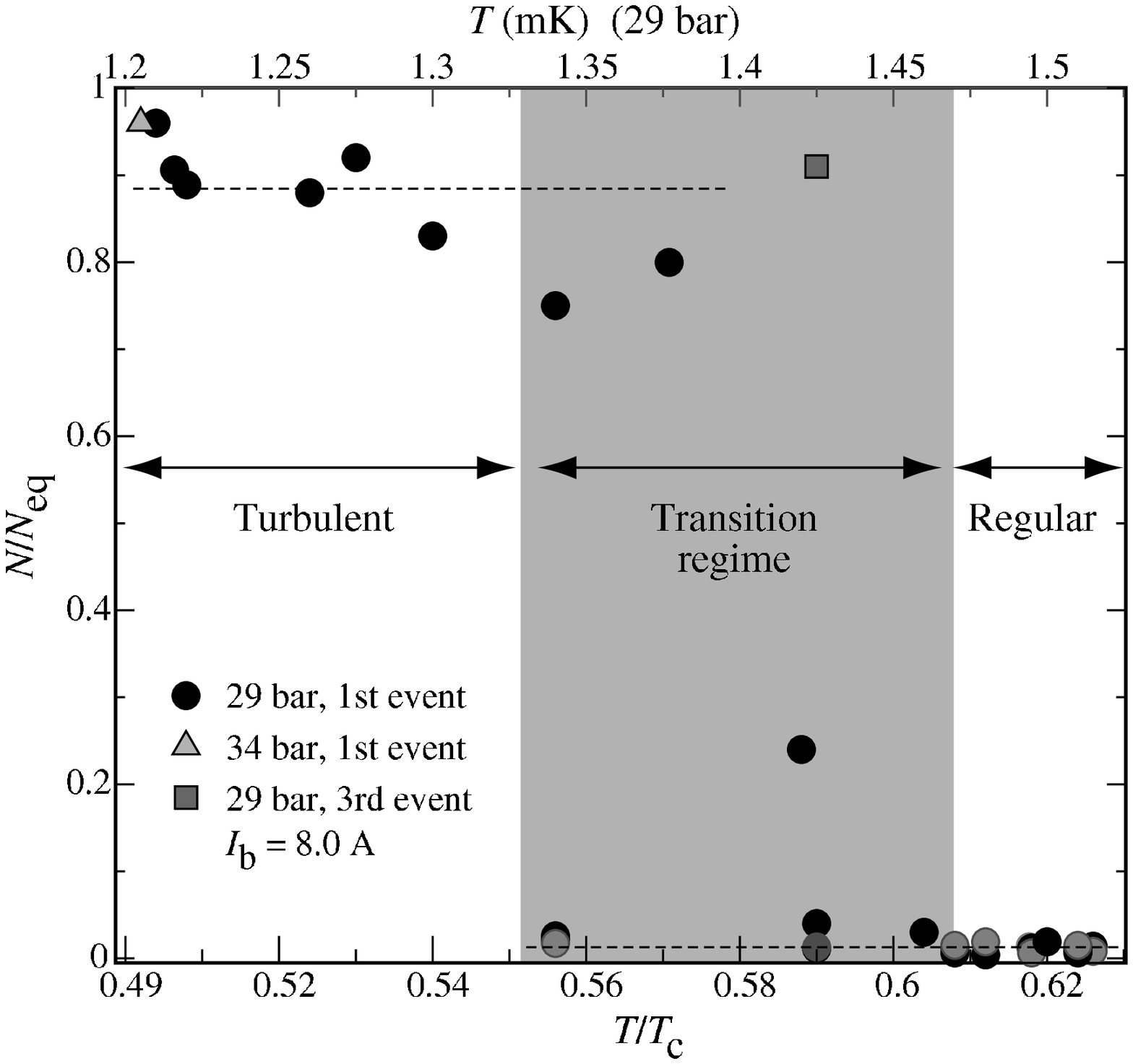}
\caption{Number of rectilinear B-phase vortex lines $N$ after KH
instability, normalized to the equilibrium number $N_{\rm eq}$ and
plotted versus temperature. At around $0.58\,T_\mathrm{c}$, a
sudden change in the number of lines is observed. At temperatures
above the transition the injection results in only a few lines,
but below the transition close to the equilibrium number is
counted. } \label{fillingratio} \vspace{-5mm}
\end{center}
\end{figure}
%%%%%%%%%%%%%%%%%%%%%%%%%%%%%%%%%%%%%%%%%%%%%%%%%%%%%%%%%%%%%%%%%%

\begin{figure}[t]
\centerline{\includegraphics[width=0.61\linewidth]{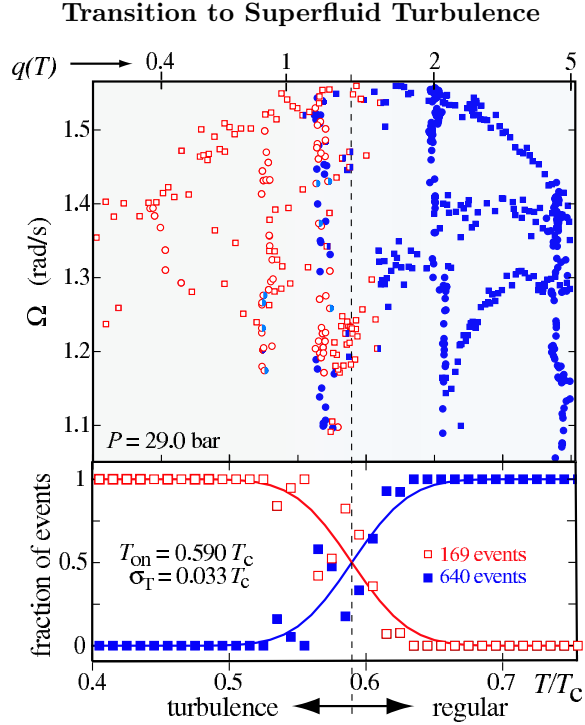}}
\medskip
\caption{Temperature -- velocity phase diagram of turbulence. Each
data point represents the result from a KH injection measurement
where $\Omega$ is increased from zero to $\Omega_{\rm c}$ at
constant temperature and the number of rectilinear vortex lines
$N$ is measured in the final state. The result is marked with an
open symbol, when the number is close to that in the equilibrium
state $N_{\rm eq}$ and when turbulence must have followed
injection. Regular vortex expansion with no increase in $N$ is
denoted with filled symbols. The vertical dashed line marks the
boundary between the two final states. The lower plot shows the
distribution of the final states in terms of gaussian fits. The
horizontal top axis gives the mutual friction parameter $q(T)$. }
\label{PDFig} \vspace{-5mm}
\end{figure}

At temperatures above $0.6\,T_{\rm c}$ the dynamics of vortex
loops injected in vortex-free flow of superfluid $^3$He-B is
regular and their number does not increase during their
time-dependent evolution to rectilinear vortex lines of the
rotating state. At lower temperatures it becomes possible for the
vortices within a bundle of vortex loops, which is injected in a
KH instability in the B phase, to start interacting turbulently.
Surprisingly this happens within a narrow temperature regime of
$0.06\,T_{\rm c}$ width centered at about $0.59\,T_{\rm c}$, as
seen in Figs.~\ref{fillingratio} and \ref{PDFig}. The consequence
from this abrupt change in the dynamics, while the vortices evolve
in the externally applied flow towards their rectilinear final
state, is a radical change in the final number of vortex lines:
While at temperatures above the transition the number of
rectilinear vortices in the final state is a few and reproduces
the distribution of vortices crossing the AB interface in one KH
instability event, at temperatures below the transition the final
state includes close to the equilibrium number of vortex
lines.\cite{Nature} These properties can be examined in great
detail, since the KH critical velocity stays well-behaved as a
function of temperature and is a continuous smooth curve across
this division line, continuing to follow the calculated
dependence. All this indicates that the transition is not related
to the properties of the KH instability itself, but arises from
the change in the dynamics of vortices moving in rapidly flowing
$^3$He-B.

%%%%%%%%%%%%%%%%%%%%%%%%%%%%%%%%%%%%%%%%%%%%%%%%%%%%%%%%%%%%%%%%%%
%%%%%%%%%%%%%%%%%%%%%%%%%%%%%%%%%%%%%%%%%%%%%%%%%%%%%%%%%%%%%%%%%%

In Fig.~\ref{PDFig} the phase diagram of regular and turbulent
vortex dynamics is shown as a function of flow velocity (vertical
scale) and temperature (horizontal scale). The vertical axis is
expressed in terms of the rotating B-phase counterflow velocity
$|{\bf v}_{\rm B} - {\bf v}_{\rm n}|=\Omega R$. In this diagram
each data point represents a KH injection measurement, obtained
with different settings of the externally controlled parameters
$T$ and $I_{\rm b}$, so that a variation as wide as possible for
the critical velocity $\Omega_{\rm c}$ is obtained. What we are
interested in here is the division in filled and open symbols:
Injection events followed by a turbulent burst are marked with
open symbols while events which lead to only a few rectilinear
lines are marked with filled symbols.

The striking conclusion from Fig.~\ref{PDFig} is that the boundary
between turbulence at low temperatures and regular dynamics at
high temperatures is vertical, {\it i.e.} it is independent of the
applied flow velocity above about 2.5\,mm/s. It will be next shown
that this transition to turbulence is driven by the dimensionless
velocity independent parameter $q(T)^{-1}$, which characterizes
the friction force acting on vortices when they move with respect
to the normal component of the liquid. It is this parameter which
divides the vortex dynamics in superfluids into a low-$1/q$ regime
with regular vortex number conserving motion and a high-$1/q$
regime where superfluid turbulence becomes possible. \vspace{-5mm}

\subsection{Superfluid Equivalent of Reynolds Number}
\label{Theory}

The velocity $\bm{v}_{\rm L}$ of an element of a vortex line is
given by
\begin{equation}
\bm{v}_{\rm L}=\bm{v}_{\rm s} +\alpha \hat{\bm{s}} \times
(\bm{v}_{\rm n}-\bm{v}_{\rm s}) -\alpha' \hat{\bm{s}} \times
[\hat{\bm{s}} \times (\bm{v}_{\rm n}-\bm{v}_{\rm s})]\,,
\label{vl}
\end{equation}
where $\hat{\bm{s}}$ is a unit vector parallel to the vortex line
element. This equation of motion depends on the dimensionless
mutual friction parameters $\alpha(T)$ and $\alpha'(T)$, which
originate from the dissipative and reactive forces acting on a
vortex when it moves with respect to the normal component. For
vortices in a fermionic system they were calculated by
Kopnin.\cite{KopninBook} For $^3$He-B they were measured over a
broad temperature range by Bevan {\it et al.}\cite{Bevan} (see
also the monograph [\onlinecite{VolovikBook}], where these
parameters are discussed in terms of the chiral anomaly).
Evidently the nature of the solutions of Eq.~(\ref{vl}) for
$\bm{v}_{\rm L}$ has to depend on the mutual friction parameters.

This property will become clearer if we form the coarse-grained
hydrodynamic equation for superfluid vorticity $ {\bm
\omega}=\nabla\times {\bf v}_{\rm s}$, by averaging over vortex
lines. This equation for the superfluid velocity can be obtained
directly from the Euler equation for inviscid liquids where,
instead of the viscous $\nabla^2{\bf v}$ term of the Navier-Stokes
equation, one then has (see review in
Ref.~[\onlinecite{soninhydro}])
\begin{eqnarray}
\frac{\partial {\bf v} }{ \partial t}+ \nabla\mu= {\bf v} \times
 {\bm \omega}~-
\alpha'({\bf v} -{\bf v}_{\rm n})\times
  {\bm \omega}+  \alpha~\hat {\bm \omega}
\times( {\bm \omega} \times({\bf v} -{\bf v}_{\rm n}) ) ~.
\label{SuperfluidHydrodynamics2}
\end{eqnarray}
Here ${\bm v}\equiv {\bm v}_{\rm s}$ is the superfluid velocity, $
{\bm \omega}=\nabla\times {\bf v}$ the superfluid vorticity, and
$\hat {\bm \omega}$ the unit vector $\hat {\bm \omega}= {\bm
\omega}/\omega$. In $^3$He-B the normal component has oil-like
viscosity, its motion is practically always laminar, and we ignore
its dynamics. In the frame where it is at rest, {\it i.e.} ${\bf
v}_{\rm n}=0$, the equation for superfluid hydrodynamics is
simplified:
\begin{equation}
\frac{\partial {\bf v} }{ \partial t}+ \nabla\mu= (1-\alpha'){\bf
v} \times
 {\bm \omega}+  \alpha~\hat{\bm\omega} \times( {\bm \omega}
\times{\bf v} ) ~. \label{SuperfluidHydrodynamics}
\end{equation}
After rescaling the time, $\tilde t=(1-\alpha')t$, one obtains an
equation which depends on a single parameter
$q=\alpha/(1-\alpha')$:
\begin{equation}
\frac{\partial {\bf v} }{ \partial \tilde t}+ \nabla \tilde\mu-
 {\bf v} \times
 {\bm \omega}=  q~\hat{\bm\omega} \times( {\bm \omega}
\times{\bf v} ) ~. \label{SuperfluidHydrodynamics3}
\end{equation}
Let us compare this equation with the Navier-Stokes equation
\begin{equation}
\frac{\partial {\bf v}}{ \partial t}+ \nabla\mu- {\bf v}\times
 {\bm \omega}=  \nu\nabla^2 {\bf v}
~. \label{NormalHydrodynamics}
\end{equation}
The inertial terms on the lhs of
Eq.~(\ref{SuperfluidHydrodynamics3}) for superfluid hydrodynamics
are the same as those on the lhs of
Eq.~(\ref{NormalHydrodynamics}) for viscous hydrodynamics. In
contrast, the dissipative term on the rhs of
Eq.~(\ref{SuperfluidHydrodynamics3}) is different from the
corresponding viscous term on the rhs of
Eq.~(\ref{NormalHydrodynamics}). This difference between the
dissipative terms in the two equations of motion is of importance.
Reynolds number, which characterizes the nature of the solutions,
is formed as the dimensional equivalent of the ratio of the
inertial and dissipative terms in the two hydrodynamic equations
(\ref{SuperfluidHydrodynamics3}) and (\ref{NormalHydrodynamics}).
For the Navier-Stokes case Eq.~(\ref{NormalHydrodynamics}) the
ratio corresponds dimensionally to the conventional Reynolds
number $Re=UD/\nu$, where $U$ is the characteristic velocity scale
and $D$ the size of the large-scale flow. In superfluid
hydrodynamics Eq.~(\ref{SuperfluidHydrodynamics3}) this ratio is
simply given by the dimensionless intrinsic parameter $1/q$. Since
$1/q$ does not depend on velocity or on the large-scale system
size, turbulence becomes possible only if $1/q$ is large enough.
In accordance to the usual characterization of the solutions of
the Navier-Stokes equation (\ref{NormalHydrodynamics}) we might
thus expect the transition to turbulence to occur at $1/q \sim 1$.
This is in agreement with the experimental phase diagram in Fig.
\ref{PDFig}.

In the above considerations we used the coarse-grained description
of superfluid vorticity, which does not take into account the
discreteness of quantized vortices. This implies, that the flow
velocity is large enough, {\it i.e.} it is essentially higher than
the Feynman critical velocity, $U\gg \kappa/D$, where $\kappa$ is
the circulation quantum. This condition is always fulfilled in
$^3$He-B experiments, in which vortices are injected via the KH
instability, since it occurs in the limit where the critical value
of $1/q$ does not depend on the flow velocity, as seen in Fig.
\ref{PDFig}. However, at much lower velocities the critical value
of $1/q$ is ultimately expected to increase.

Furthermore, the coarse-grained description of quantized vorticity
for estimating the threshold value for $1/q$ assumes a
sufficiently large number and density of injected vortices. The
number of vortices injected in the KH instability is on average 10
(in a distribution which ranges from 3 to 30). They are packed
close to each other since they originate from the same growing,
but over-damped, ripplon corrugation (Fig.
\ref{KelvinInstabilityFig}). However, in other types of injection
experiments the number and density of injected vortices can be
less,\cite{Precursor,Solntsev} and the critical value of $1/q$
increases. This is similar to what occurs in viscous pipe
flow,\cite{Mullin} where the critical Reynolds number $Re$ for the
onset of turbulence depends on the magnitude of the perturbation.

The transition from regular to turbulent dynamics as a function of
the mutual friction parameter $1/q$ is a new phenomenon. It has
not been observed in $^4$He-II, where $1/q$ is practically always
large, so that a transition is expected only a few tens of
microkelvins below $T_{\lambda}$ where $\rho_{\rm s}$ is
vanishingly small, the coherence length $\xi(T)$ diverges,
critical velocities approach zero, and vortex dynamics enters a
regime where little is known. In $^3$He-B the transition at
$1/q\sim 1$ is in the middle of the experimentally accessible
temperature range and can be observed in one single experiment by
scanning temperature from the superconductor-like dynamics (with
no pinning, but high vortex damping) at high temperatures to
superfluid $^4$He-like turbulence at low temperatures.

These features demonstrate that vortex dynamics in superfluids
takes varied forms and that the traditional $^4$He-like behavior
is just one extreme example. The opposite extreme is the strongly
anisotropic superfluid $^3$He-A where $1/q$ is practically always
in the range of regular dynamics and the unusually low
temperatures, which would be required for conventional superfluid
turbulence, are not experimentally realistic at this time.
Instead, in $^3$He-A it is transitions in the structure of the
vortex and in the global order parameter texture which provide a
means to adjust to faster dynamics. An example of this is the
dynamically driven transition from vortex lines to sheets in
applied flow which is periodically alternating in
direction\cite{LinesToSheets} and the dependence of the critical
velocity on the global order parameter texture\cite{CritVel}.
\vspace{-5mm}

\subsection{Turbulence in Uniform Rotation} \label{TurbulentBurst}

%%%%%%%%%%%%%%%%%%%%%%%%%%%%%%%%%%%%%%%%%%%%%%%%%%%%%%%%%%%%%%%%%%%%
\begin{figure}[t]
\centerline{\includegraphics[width=\linewidth]{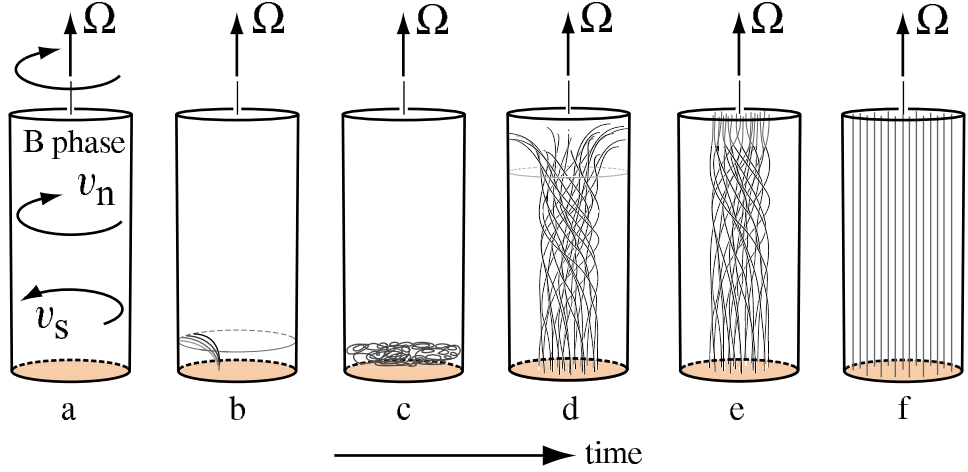}}
\medskip
\caption{Temporal evolution of vorticity in a rotating column,
shown schematically (in the rotating frame) following the
injection of seed vortices across the AB interface in a KH
instability (at $T>0.40\,T_{\rm c}$). ({\bf a}) Vortex-free
initial state above AB interface. ({\bf b}) A tight bundle of
$\sim 10$ vortex loops is ejected across the AB interface in a KH
instability. ({\bf c}) A brief burst of turbulence spreads across
the entire cross section next to the AB interface. Close to the
equilibrium number of independent vortices is created and
immediately polarized. ({\bf d}) The vortices expand in spiral
motion along the column with a sharp front towards the vortex-free
section. ({\bf e}) When the front reaches the top end plate, the
twisted vortex state continues relaxing to rectilinear lines. If
the twist is not too tight, then it is  removed by the slow
diffusive motion of the vortex ends on both end plates. ({\bf f})
Final stable equilibrium vortex state with rectilinear lines.}
\label{TurbulenceInRotation} \vspace{-5mm}
\end{figure}
%%%%%%%%%%%%%%%%%%%%%%%%%%%%%%%%%%%%%%%%%%%%%%%%%%%%%%%%%%%%%%%%%%%%

So far measurements on the transition to turbulence have been
performed in uniformly rotating flow of $^3$He-B. The long
smooth-walled sample container was prepared from fused quartz with
an aspect ratio (length to diameter) of almost 20. The sample can
be examined in single phase configuration filled with only B
phase, when the barrier magnet is not charged. In the opposite
case the barrier field stabilizes over the central section of
sample length a narrow layer of A phase which acts as a vortex
barrier. The two end sections of equal length with B phase
represent then two independent samples, whose lengths are roughly
half of that of the single phase sample. In this way one can
examine the influence of the length of the cylinder on the
evolution and propagation of vortices in the rotating column.
Another bonus is the possibility to study two B-phase samples in
parallel when the A-phase barrier layer is present. This feature
is of importance since the critical velocity of the sample
container is an unknown predicament. The precondition for KH
injection measurements is a high critical velocity. This restricts
measurements to an open cylindrical volume with smooth clean
walls, no internal measuring probes, and isolated from the rest of
the liquid $^3$He volume by a small orifice in the center of one
of the flat end plates. The number of rectilinear vortices in
these two B phase sections is monitored with non-invasive NMR
absorption measurements, by means of NMR detector coils mounted
outside around both ends of the long quartz cylinder.

Rotation provides a strong polarization for the vorticity to be
oriented along the rotation axis. It is known from studies of
turbulence in rotating viscous fluids that turbulent disorder then
tends to become limited to the transverse plane. A similar
phenomenon occurs in thermal counterflow of rotating $^4$He-II, if
the thermal current is imposed parallel to the rotation axis so
that it counteracts the polarization from rotation.\cite{Tsubota}
Thus turbulence in effect tends to become 2-dimensional in
rotation. In the case of KH injection turbulence is limited to a
brief burst which takes place in the cross section of the cylinder
next to the AB interface where the injection occurs. In this short
burst the equilibrium number of vortices is created and, because
of rapid polarization, the free energy drops instantaneously
within this cross section from the maximum close to the minimum
value. In Fig.~\ref{TurbulenceInRotation} the ensuing evolution of
the vorticity is shown schematically. It consists of the winding
cork-screw-like propagation of the vorticity along the rotating
column, and the final straightening of the helically twisted
vortices towards rectilinear lines of the stable equilibrium
vortex state.\cite{TwistedState}

The spiralling vortex motion in the rotating column can be seen to
emerge from Eq.~(\ref{vl}). Let us write the expression for the
velocity of the end point of a vortex, where it connects in
perpendicular orientation to the cylindrical side wall of the
sample container. In the vortex front this end point practically
resides in vortex-free counterflow with ${\bm v}_{\rm n} = \Omega
R {\hat {\bm \phi}}$ and ${\bm v}_{\rm s} = 0$, so that its
velocity is
\begin{equation} {\bm v}_{\rm L} = -(1-\alpha^{\prime})\Omega R
{\hat {\bm \phi}} + \alpha \Omega R {\hat {\bm z}}~,\end{equation}
which includes two components. The first component enforces
azimuthal motion which differs by the fraction $1-\alpha^{\prime}$
from the velocity $-\Omega R$ at which the superfluid component
travels with respect to the cylinder wall. The second component
moving with the velocity $\alpha \Omega R$ corresponds to the
longitudinal expansion of the vortex end along the column.

With decreasing temperature the longitudinal motion slows down and
ultimately becomes exceedingly slow. Since the KH instability
occurs at a well-defined critical velocity, it can be started as a
triggered event and the longitudinal expansion velocity can be
measured by timing the time delay from the trigger to the moment
when the vortex front reaches a detector coil.\cite{Turbulence} At
higher temperatures above $0.40\,T_{\rm c}$ the longitudinal
velocity of the front is found to be close to $\alpha \Omega
R$.\cite{TimeOfFlight} Thus the above scenario explains
qualitatively the formation of the twisted vortex state behind the
vortex front. The twisted state corresponds to dynamic equilibrium
in a `force-free configuration'.\cite{TwistedState} This means
that the residual superfluid velocity created by the twist of the
vortices, which has both radial and longitudinal components, is
strictly oriented along the vortex cores. With decreasing
temperature the vortex front becomes sharper, the longitudinal
expansion is slowed down, the twist in the trailing vortex bundle
is wound ever tighter, and the residual superflow velocity along
the vortices increases. Ultimately the flow along the vortices is
expected to make them unstable with respect to the
Glaberson-Donnelly Kelvin-wave instability and inter-vortex
reconnections will become frequent behind the vortex
front.\cite{KelvinWaveInstability} Such reconnections help to
remove the twist and to speed up the relaxation of the twisted
state towards the stable equilibrium vortex state with rectilinear
lines (Fig.~\ref{TurbulenceInRotation}). \vspace{-5mm}

\section{CONCLUSION}
\label{ConclusionSec}

Although the two helium superfluids, $^4$He-II and $^3$He-B, were
expected to obey similar hydrodynamics, they differ somewhat in
the actual values of their hydrodynamic properties. Rather
surprisingly, these small differences have led to new insight in
superfluid dynamics during recent years. An example is the
transition between turbulent and regular vortex dynamics as a
function of mutual friction dissipation $\alpha(T)$. In the
superfluid literature a `transition to turbulence' has so far been
associated exclusively with the transition in $^4$He-II as a
function of the applied flow velocity.\cite{Skrbek} These
velocities, at which vortices typically become mobile in
$^4$He-II, are very low compared to the intrinsic critical
velocity of the bulk superfluid. In contrast, the mutual friction
driven transition to turbulence at high flow velocities is
velocity independent. Among the various coherent quantum systems
$^3$He-B is unique in that $\alpha(T)$ spans the range where $1/q
= (1-\alpha^{\prime})/\alpha$ passes through the transition at
$1/q \sim 1$. A similar situation might become available in atomic
Bose-Einstein condensates, where mutual friction\cite{BEC-mf} has
also been predicted to include the regime $1/q \sim 1$ and where
turbulence might soon be experimentally
realized\cite{BEC-turbulence}.

A second advantage in using $^3$He-B for vortex dynamics studies
is the high viscosity of its normal component. This guarantees
that the normal component is practically always in a state of
laminar flow and provides a well-behaved reference frame for
superfluid dynamics. This simple situation is quite unlike that in
$^4$He-II, where the coupled dynamics of the superfluid and normal
components have to be included from the start. Finally we note
that owing to the much larger superfluid coherence length $\xi(T)
> 10\,$nm, which determines the length scale for the radius of
the vortex core in isotropic He superfluids, critical velocities
can be controlled more efficiently in $^3$He-B. This allows new
types of experimental studies. One example is the externally
controlled introduction of seed vortices in vortex-free flow which
made it possible to identify the mutual-friction-dependent
transition to turbulence. The superfluid KH instability of the AB
interface in a two-phase sample of $^3$He superfluids is one of
the unique and most reproducible of such injection methods. At
present time this is the method by which a direct transition to
bulk turbulence is believed to start at the lowest possible value
of $1/q$.

{\bf Acknowledgements:} This work is supported in part by the EU
Trans-national Access Programme FP6 ($\#$RITA-CT-2003-505313), by
the Academy of Finland via its 2006 grant for visitors from
Russia, by the Russian Ministry of Education and Science (Leading
Scientific School grant $\#$1157.2006.2), and by the European
Science Foundation COSLAB Program. 
\end{document}